\documentstyle[11pt,printmor00,epsfig]{article}

\def\bq{\begin{quote}}
\def\eq{\end{quote}}

\def\ep{\epsilon}

\def\beq{\begin{equation}}
\def\eeq{\end{equation}}

\def\bea{\begin{eqnarray}}
\def\eea{\end{eqnarray}}

\begin{document}
\vspace*{4cm}
\title{A SEE-SAW MECHANISM FOR LARGE NEUTRINO MIXING \\FROM SMALL 
QUARK AND LEPTON MIXINGS}

\author{ ISABELLA MASINA }

\address{Universit\`a di Padova and I.N.F.N., Sez. di Padova, 
Via Marzolo 8, Padua, Italy}

\maketitle
%\centerline{\bf Abstract}
\abstracts{\centerline{\bf Abstract} 
I introduce and sketch the main features of those
see-saw models \cite{afm1} 
where a large atmospheric mixing can be achieved starting from 
nearly diagonal matrices for charged leptons, Dirac neutrinos 
and Majorana right-handed neutrinos. It turns out that these models can
be realized in Grand Unified Theories and they are well compatible with
the related phenomenology of fermion masses and mixings.
}

\section{Introduction}

Recent impressive results from Superkamiokande 
\cite{SK} have cofirmed that the atmospheric neutrino anomaly 
can be successfully interpreted in terms of neutrino oscillations. 
Also the solar neutrino deficit, observed by several experiments 
\cite{solexp}, is probably an indication of a different sort 
of neutrino oscillations. 
Since neutrino oscillations imply neutrino masses, 
one is forced to look for 
viable extensions of the Standard Model. 
In doing this, it is worth
to stress that the extreme smallness of neutrino 
masses in comparison with quark and charged lepton masses 
seems to point in favour of a different nature of the former, 
maybe linked to lepton-number violation. 
Experimental facts on neutrino masses could then provide 
an indication on the very large energy scale where 
lepton-number is violated.
Grand Unified Theories (GUTs) are certainly a very attractive 
framework where neutrino masses can be analyzed, because they  
predict - besides, of course, 
unification of the three gauge coupling constants -  
lepton and baryon-number violation.
Since in GUTs all fermion masses 
are related, neutrino masses and mixings could also provide 
an insight on the mechanism for the generation 
of charged fermion masses. 
In particular the observation 
of a nearly maximal mixing angle for 
$\nu_{\mu}\rightarrow \nu_{\tau}$ is particularly impressive. 
At present solar neutrino mixings can be either large 
or very small, depending on which particular 
solution will eventually be established by the data. 
Large mixings in the neutrino sector are very interesting 
because a first guess was in favour of small mixings, in analogy to 
what is observed for left mixings in the quark sector. 
If confirmed, single or double maximal mixings can provide 
an important hint on the mechanisms that generate neutrino masses.

In the context of GUTs, many theoretical descriptions 
of large neutrino mixing(s) have been discussed (see Ref.~4 
for a review). 
In most models large mixings are already present at the 
level of Dirac and/or Majorana matrices for neutrinos. 
Instead, here I discuss the interesting class of
models where large, possibly maximal, neutrino mixings 
are generated by the see-saw mechanism starting from 
nearly diagonal Majorana and Dirac matrices for 
neutrinos, without fine-tuning or stretching 
small parameters into becoming large. 
A more complete discussion of the subject can be found in Ref.~1.

With neutrino masses settled, observation of proton decay will be
the next decisive challenge remained to support or eventually 
put in crisis GUTs. In fact, SuperKamiokande \cite{SKp} is giving lower 
bounds on the proton life-time which, for certain decay modes, yet exclude 
part of the range generally predicted by GUTs. Then it seems interesting
to see if it is possible to construct a simple but realistic GUT model 
\cite{afmp} which
not only correctly reproduces the informations on neutrino masses and the
actual bounds on proton decay, but also overcome the typical problem 
of these theories, that is the doublet-triplet splitting problem,
without destroying gauge coupling unification.

\section{Starting Assumptions}

Since the experimental status of neutrino oscillations 
is still very preliminary, one has to make a
number of assumptions on how the data will finally look like. 
Here I assume that only two distinct oscillation
frequencies exist, the largest being associated 
with atmospheric neutrinos and the smallest with solar neutrinos. 
I assume that the hint of an additional frequency 
from the LSND experiment \cite{LSND} will disappear, thus  
avoiding the introduction of new sterile neutrino species. 
Dealing with only the three known
species of light neutrinos, the atmospheric neutrino 
oscillations are interpreted as nearly maximal 
$\nu_{\mu}\rightarrow \nu_{\tau}$ oscillations 
while the solar neutrino oscillations correspond to the 
disappearance of $\nu_e$ into nearly equal fractions of 
$\nu_{\mu}$ and $\nu_{\tau}$. One has to 
be open minded to all the three most likely 
solutions for solar neutrino oscillations \cite{Lisi}: 
the two MSW solutions
with small (SA) or large (LA) mixing angle, 
or the vacuum oscillation solution (VO).
Assuming only two frequencies, given by
\beq
~~~~\Delta_{sun}\propto m^2_2-m^2_1~~~,~~
\Delta_{atm}\propto m^2_3-m^2_{1,2}~~~,
\eeq
there are two extreme possibilities for the mass eigenvalues:
\beq
    {\rm A}~ : ~ m_3 >> m_{2,1}~~~~~~~~~{\rm B}~ : ~ m_1 \sim m_2 \sim m_3 ~~.
\eeq
Configuration B imply a very precise near degeneracy 
of squared masses: it would need a relative splitting 
$|\Delta m/m|\sim \Delta m^2_{atm}/2m^2\sim 10^{-3}$--$10^{-4}$ 
and a much smaller one for solar neutrinos, 
especially if explained by vacuum oscillations: 
$|\Delta m/m|\sim 10^{-10}$--$10^{-11}$. 
Foreseeing a GUT framework, it is reasonable to assume 
that the Dirac neutrino matrix has a strongly hierarchical structure, 
as is the case for charged fermions. So, it seems quite
implausible that, starting from hierarchical Dirac matrices, 
one end up via the see-saw mechanism into a
nearly perfect degeneracy of squared masses. 
As a consequence, here I will focus on models of type A 
with large effective light neutrino mass splittings and large mixings.

\section{A $2\times 2$ Example }

Reconciling large splittings with large mixing would seem difficult. 
Indeed, one could guess that, in analogy to what is observed for 
quarks, large splittings correspond to small
mixings because only close-by states are strongly mixed. 
At the contrary, via the see-saw mechanism \cite{ss}, there are two 
particularly simple ways in which this can be realized.

Without loss of generality, leaving apart for the moment the eventual presence
of flavor symmetries, one can go to the basis where both the charged 
lepton Dirac mass matrix $m_D^l$ and the Majorana matrix $M$ for the 
right-handed neutrinos are diagonal.
For simplicity, let's start assuming that the role of the first generation 
is not crucial in the mechanism for the generation of neutrino masses, 
so than one can, with good approximation, work in the 2 by 2 case 
(in the next section I will however relax this condition). 
If one writes $m_D$ (defined by $\bar R m_D L$) and $M$ in the 
most general way:
\beq
m_D=v\left[\matrix{a&b \cr c&1}\right],
~~~~~ M=\left[\matrix{M_2&0\cr 0&M_3}\right]~~~~~,
\label{mdM}
\eeq
where $v$ is a vacuum expectation value, $a,b$ and $c$ are Yukawa 
couplings, then, via the see-saw, one obtains:
\beq
m_{\nu}=\frac{v^{2}}{M_3}
\left[\matrix{\frac{a^{2}}{M_2}+\frac{c^{2}}{M_3}&
\frac{a~b}{M_2}+\frac{c}{M_3} 
\cr \frac{a~b}{M_2}+\frac{c} {M_3}&\frac{b^{2}}{M_2}+\frac{1}{M_3} }
\right]~~~~~.
\label{mnM}
\eeq

The request of large splittings among the light neutrino's eigenvalues
is equivalent to demanding that the determinant of the previous matrix is
much smaller than its trace. 
It is then possibile to see at first sight that two very natural cases 
arise, respectively when the terms with $M_3$ or $M_2$ at the denominator
are dominant.

One simple example of the first case is realized if $M_2 \sim M_3$ and 
$a,b \ll 1$. In order to have a large splitting, one must 
have $c\sim 1$, that is the right-handed neutrino of the third 
generation couples with the same strenght \cite{King} to left-handed 
$\nu_{\mu}$ and $\nu_{\tau}$. 
The heaviest mass for light neutrinos results
$m_{3}\sim v^{2}/M_3$. Since in the hierarchical case, the data from 
SuperKamiokande suggest $m_{3}\sim 0.05 eV$, if one assumes that $v$ is
a typical weak scale, namely 250 GeV, then $M_3\sim 10^{15} GeV$, just
the order of a GUT scale.
It is worth to stress that this first mechanism is based on asymmetric Dirac 
matrices with, in the case of the example, a large
left-handed mixing already present in the Dirac matrix. 
It has been observed \cite{af,asymm} that in SU(5)
GUT left-handed mixings for leptons tend to correspond to right-handed 
mixings for $d$ quarks (in a basis where $u$ quarks are
diagonal). Since large right-handed mixings for quarks are not in 
contrast with experiment, viable GUT models that correctly reproduce 
the data on fermion masses and mixings can be constructed following 
this mechanism.

An alternative possibility \cite{afm1} is to have the dominance 
of the terms with
$M_2$ at the denominator. 
This is achieved for any $c<1$ if 
$a^{2},b^{2}> M_2/M_3$. 
The request for large splitting is then equivalent to
require also $a\sim b$. 
Now it is the second generation right-handed neutrino 
which is particularly light and which couples with the same strenght
\cite{King} 
to left-handed $\nu_{\mu}$ and $\nu_{\tau}$. 
In order to be more specific, consider one particular example with
symmetric matrices. 
These matrices are interesting because, for instance,
one could want to preserve left-right symmetry at the GUT scale. 
Then, the observed smallness of left-handed mixings for quarks would
also demand small right-handed mixings. 
Starting from 

\beq
 m_D=v
\left[\matrix{\ep&x\ep\cr x\ep&1    }
\right],~~~~~ M^{-1}=\frac{1}{M_3}
\left[\matrix{r_2&0\cr 0&1    }
\right]~~~~~,
\label{mdM}
\eeq
where $\ep$ is a small number, $x$ is of O(1) and $r_2\equiv M_3/M_2$, 
then, via the see-saw, 
it is sufficient that $\ep^2 r_2\gg 1$ in order to have approximately:

\beq
m_{\nu}=\frac{v^2}{M_3} 
\epsilon^2 r_2\left[\matrix{1&x\cr x&x^2}
\right]~~~~~.
\label{mnp1}
\eeq

The determinant is naturally vanishing so that the
mass eigenvalues are widely split and for $x\sim 1$ the mixing is nearly
maximal. It is exactly maximal if $x=1$.
The see-saw mechanism has created 
{\it large mixing from almost nothing}: all
relevant matrices entering the see-saw mechanism are nearly diagonal
\cite{afm1,Smir}, 
that is they are diagonalized by transformations that go into the 
identity in the limit of vanishing $\ep$. 
Clearly, the crucial
factorization of the small parameter $\epsilon^2$ only arises if 
the light Majorana eigenvalue is coupled to $\nu_\mu$ and
$\nu_\tau$ with comparable strength, that is $x\sim 1$.
An interesting feature of this second case, in connection with a possible
realization within a GUT scheme, is that it requires 
$M_3>v^2/m_3$, so that one can push $M_3$, the scale of 
lepton-number violation, beyond the GUT scale. 
This is desirable because, for instance, this is expected in SU(5) 
if right-handed neutrinos are present and also in the breaking of 
SO(10) to SU(5).

Summarizing, the second case require a peculiar hierarchy in the 
Majorana eigenvalues in order to work, but it has however the good 
characteristic that it can be realized even with nearly diagonal matrices.

\section{Generatization to the $3\times 3$ Case}

It is straightforward to extend the previous model to the 3 by 3 case. 
One simple class of examples with symmetric mass matrices is the
following one.
Starting from
\beq
 m_D=v
\left[\matrix{\epsilon''&\epsilon'& y~\epsilon'\cr
\epsilon'&\epsilon &x~\epsilon\cr y~\epsilon' &x~\epsilon &1      }
\right],~~~~~ M^{-1}=\frac{1}{\Lambda}
\left[\matrix{ r_1&0&0\cr 0& r_2&0\cr 0&0& r_3    }
\right]~~~~~, 
\label{mdM3}
\eeq 
where, unless otherwise stated, $x$ and  $y$ are O(1); $\epsilon$, 
$\epsilon'$ and $\epsilon''$
are independent small numbers and $r_i\equiv M_3/M_i$. One
expects $\epsilon''\ll\epsilon'\ll\epsilon\ll 1$ and, perhaps, also $r_1\gg r_2\gg r_3=1$,
if the hierarchy for right-handed
neutrinos follows the same pattern as for known fermions. Depending on the relative size of the ratios
$r_i/r_j$,
$\epsilon/\epsilon'$ and $\epsilon'/\epsilon''$, it is possible to have models with dominance of any of the
$r_{1,2}$.
For example, setting $x=1$ (keeping $y$ of O(1)) and
assuming $r_2\epsilon^2\gg r_1\epsilon'^2,r_3$, together with  $r_2 \epsilon'^2\gg r_1 \epsilon''^2$ and
$r_2\epsilon\gg r_1 \epsilon''$, with good accuracy we obtain:
\beq  
m_{\nu}=
\frac{v^2}{\Lambda}r_2\epsilon^2
\left[\matrix{
\frac{\epsilon'^2}{\epsilon^2}
&\frac{\epsilon'}{\epsilon}
&\frac{\epsilon'}{\epsilon}
\cr
\frac{\epsilon'}{\epsilon}
&1+\frac{r_1\epsilon'^2}{r_2\epsilon^2}&
1
\cr
\frac{\epsilon'}{\epsilon}
&1
&1+\frac{r_3}{r_2\epsilon^2}    }
\right]~~~~~.
\label{mnFI}
\eeq

Since the subdeterminant of the 23 block is vanishing, 
the eigenvalues are widely split. Having set $x=1$
the atmospheric
neutrino mixing is nearly maximal.
The solar neutrino mixing is instead generically
small in these models, being proportional to $\epsilon'/\epsilon$. 
Thus the SA-MSW solution is obtained. It is
easy to find
set of parameter values that lead to an acceptable phenomenology 
within these solutions.
As an illustrative example take:
\beq
\epsilon\sim \lambda^4~~,~~~~\epsilon'\sim \lambda^6~~,~~~~
\epsilon''\sim\lambda^{12}~~,
~~~~r_1\sim\lambda^{-12}~~,~~~~r_2\sim\lambda^{-9}~~,
\eeq
where $\lambda\sim\sin\theta_C$, $\theta_C$ being the Cabibbo angle. 
The neutrino mass matrices 
than become
\beq  
m_D=v
\left[\matrix{\lambda^{12}&\lambda^6&\lambda^6\cr
\lambda^6&\lambda^4&\lambda^4\cr \lambda^6&\lambda^4&1      }
\right]~~,~~~~~~M=\Lambda
\left[\matrix{\lambda^{12}&0&0\cr
0&\lambda^9&0\cr 0&0&1      }
\right]
\label{ex1}
\eeq
and, in units of $v^2/\Lambda$, we obtain: $~~m_3\sim 1/\lambda~~,
~~~~~m_2\sim 1~~,~~~~~m_1\sim\lambda^4~~$.
The solar mixing angle $\theta_{12}$ is
of order $\lambda^2$, suitable to the SA-MSW solution. Also 
$\theta_{13}\sim\lambda^2$.

Models based on symmetric matrices are
directly compatible with left-right symmetry and therefore are 
naturally linked with SO(10). This is to be
confronted with
models that have large right-handed mixings for quarks, which, 
in SU(5), can be naturally translated into large
left-handed
mixings for leptons. In this connection it is interesting to 
observe that the proposed textures for the
neutrino Dirac
matrix can also work for up and down quarks. For example, the matrices
\beq
 m_D^u\propto
\left[\matrix{0&\lambda^6&\lambda^6\cr
\lambda^6&\lambda^4&\lambda^4\cr \lambda^6&\lambda^4&1      }
\right]~~,~~~~~~ m_D^d\propto
\left[\matrix{0&\lambda^3&\lambda^3\cr
\lambda^3&\lambda^2&\lambda^2\cr \lambda^3&\lambda^2&1      }
\right]~~,
\label{mud}
\eeq
where for each entry the order of magnitude is specified in terms 
of $\lambda\sim \sin{\theta_C}$, lead to
acceptable mass
matrices and mixings. In fact $m_u:m_c:m_t=\lambda^8:\lambda^4:1$ 
and $m_d:m_s:m_b=\lambda^4:\lambda^2:1$.
The $V_{CKM}$ matrix receives a
dominant contribution from the down sector in that the up sector 
angles are much smaller than the down sector
ones. The same kind of texture can also be adopted in the charged 
lepton sector.

\section{Bimixing}

The solar mixing angle is generically small in the class of models 
explicitly discussed above. However,
small mixing angles in the Dirac and Majorana neutrino mass matrices 
do not exclude a large solar mixing angle.
For instance, this is generated from the asymmetric, but nearly
diagonal mass matrices:
\beq
m_D=v
\left[\matrix{\lambda^6&\lambda^6&0\cr
0&\lambda^4&\lambda^4\cr 0&0&1      } 
\right]~~,~~~~~~M=\Lambda
\left[\matrix{\lambda^{12}&0&0\cr
0&\lambda^{10}&0\cr 0&0&1      } 
\right]~~~~~.
\label{ex22}
\eeq
They give rise to a light neutrino mass matrix of the kind:
\beq 
m_\nu=
\left[
\matrix{
\lambda^2&\lambda^2&0\cr
\lambda^2&1&1\cr
0&1&1}\right]{v^2\over \lambda^2\Lambda}~~~~~,
\label{mnu3}
\eeq 
which is diagonalized by large $\theta_{12}$ and $\theta_{23}$ and 
small $\theta_{13}$.
The mass hierarchy is suitable to the large angle MSW solution.

\section{Outlook and Conclusions}

Other comments about this mechanism are contained in Ref.~1.
For instance here we show that:

i) the results obtained are stable under renormalization from the high 
energy scale where the mass matrices 
are produced down to the electroweak scale; in fact, if light neutrino 
masses are hierarchical, one always expects renormalization effects
to be negligible \cite{renorm}; 

ii) it is possible to construct specific realizations of
the mechanism sketched here, e.g. in the context of SU(5) 
$\times$ broken horizontal flavour symmetries.

Summarizing, in most models \cite{af,asymm,largemix} 
that describe neutrino
oscillations with nearly maximal mixings, there appear
large mixings in at least one of the matrices $m_D$, $m_D^l$, $M$ 
(i.e. the neutrino and charged-lepton
Dirac matrices and the right-handed Majorana matrix). 
In this contribution I have discussed the peculiar possibility
that large neutrino mixing is only produced by the see-saw 
mechanism starting from all nearly diagonal
matrices.
Although this possibility is certainly rather special, 
models of this sort can 
be constructed without an unrealistic amount of fine-tuning 
and are well compatible with grand unification
ideas and the related phenomenology for quark and lepton masses.

\section*{Acknowledgments}
Many thanks go to the organizers for the stimulating 
and relaxed atmosphere of this conference, held in such beautiful
and inspiring surroundings. 
It is pleasure to thank Guido Altarelli and 
Ferruccio Feruglio for the enjoyable collaboration 
on which this talk is based.

\end{document}